 \documentclass[aps,pra,preprint,groupedaddress,amsmath,amssymb,showpacs]{revtex4}

\usepackage{graphicx}
\usepackage{amsmath}
\usepackage{dcolumn}
\usepackage{bm}

\newcommand{\im}{\mathrm{i}}

\newcommand{\ra}{\rangle}
\newcommand{\la}{\langle}

\begin{document}

\title{Depolarizing power and polarization entropy
of light scattering media: experiment and theory}

\author{Graciana Puentes}
\author{Dirk Voigt}
\author{Andrea Aiello}
\author{J.P. Woerdman}
\affiliation{Huygens Laboratory, Leiden University, P.O. Box 9504,
2300 RA Leiden, The Netherlands,
e-mail:~graciana@molphys.leidenuniv.nl}
\date{\today}
\begin{abstract}

We experimentally investigate the depolarizing power and the
polarization entropy of a broad class of scattering optical media.
By means of polarization tomography, these quantities  are derived
from an effective Mueller matrix, which is introduced through a
formal description of the multi-mode detection scheme we use, as
recently proposed by Aiello and Woerdman (arXiv:quant-ph/0407234).
This proposal emphasized an intriguing universality in the
polarization aspects of classical as well as quantum light
scattering; in this contribution  we  demonstrate experimentally
that this universality is obeyed by a surprisingly wide class of
depolarizing media. This, in turn, provides the experimentalist
with a useful characterization of the polarization properties of
any scattering media, as well as a universal criterion for the
validity of the measured data.

\end{abstract}


\maketitle

\section{Introduction}

Characterization of optically transparent media with
polarized-light scattering methods, is important for communication
technology, industrial and medical applications \cite{DOP}.  When
polarized light is incident on an optically random medium it
suffers multiple scattering and, as a result, it may emerge partly
or completely depolarized. The amount of depolarization can be
quantified by calculating
 either the entropy ($E_F$)  or
 the degree of polarization  ($P_F$) of the scattered field \cite{KligerBook}.
It is simple to show that the field quantities  $E_F$ and $P_F$
are related by a single-valued function: $E_F(P_F)$. For example
polarized light ($P_F = 1$) has $E_F =0$ while partially polarized
light ($0 \leq P_F < 1$) has $1 \geq  E_F > 0$. When the incident
beam is polarized  and the output beam is partially polarized, the
medium is said to be depolarizing. An average measure of the
depolarizing power of the medium is given by the so called index
of depolarization ($D_M$) \cite{Gil86}. Non-depolarizing media are
characterized by $D_M=1$, while depolarizing media have $0 \leq
D_M<1$. A depolarizing scattering process is always accompanied by
an increase of the entropy of the light, the increase being due to
the interaction of the field with the medium. An average measure
of the entropy that a given random medium can add to the entropy
of the incident light beam, is given by the {polarization entropy}
$E_M$ \cite{LeRoy}. Non-depolarizing media are characterized by
$E_M = 0 $, while for depolarizing media  $0< E_M \leq 1$. As the
field quantities $E_F$ and $P_F$ are related to each other, so are
the medium quantities $E_M$ and $D_M$.  In a previous paper
\cite{aiello2} we showed the existence of a universal relation
$E_M(D_M)$ between the polarization entropy $E_M$ and the index of
depolarization  $D_M$ valid for any scattering medium. More
specifically, $E_M$ is related to $D_M$ by a multi-valued
function, which covers the complete regime from zero to total
depolarization. This universal relation provides a simple
characterization of the polarization properties of \emph{any}
medium, as well as a consistency check for the experimentally
measured Mueller matrices. We emphasize that the results found in
\cite{aiello2} apply both to classical and quantum scattering
processes, and might therefore become relevant for quantum
communication optical applications, where depolarization
is due to the loss of quantum coherence \cite{gisin}.\\
In this contribution, we present an experimental study of the
depolarizing properties of a large set of scattering media,
ranging from milk to multi-mode optical fibers. The results
confirm the theoretical predictions for the bounds of the
multi-valued function $E_{M}(D_{M})$. The manuscript is divided as
follows: in Section II we review the Mueller-Stokes formalism and
show the differences between deterministic (non-depolarizing) and
non-deterministic (depolarizing) scattering media. We also discuss
the statistical nature of a depolarizing process resulting from
the average (either spatial or temporal) performed in a multi-mode
detection scheme. Furthermore, in order to describe  the
transverse spatial average present in our multi-mode detection
set-up, we formally introduce the concept of an effective Mueller
matrix ($M_{\rm{eff}}$). In Section III we describe the
experimental scheme for polarization tomography that was used to
characterize the different scattering samples.  These can  be
divided into two categories: (a) non-stationary (samples which
fluctuate during the measurement time) and (b) stationary (samples
which do not fluctuate). We then show the experimental results
obtained for these samples followed by a brief discussion of the
interesting  structures in the $(E_{M},D_{M})$ plane, that were
revealed by the experiments. Finally, in Section IV we draw our
conclusions.
\section{Depolarizing and non-depolarizing media}
In the Introduction we stressed the fact that passive optical
systems may be grouped in two broad classes: depolarizing and
non-depolarizing systems. To the first class belong all media
which decrease the degree of polarization $P_F$ of the impinging
light, while to the second one belong all media which do not
decrease $P_F$. In this Section we want to make the discussion
more quantitative by using the Mueller-Stokes formalism which is
widely used for the description of the polarization state of light
beams.
\subsection{Mueller-Stokes formalism}
Consider a  quasi-monochromatic beam of light of mean angular
frequency $\omega$ \cite{Wolf03}. Let us denote with $x,y,z$ the
axes of a Cartesian coordinate system, with the $z$-axis along the
direction of propagation of the beam whose
angular spread around $z$ is assumed to be small enough to satisfy
the \emph{paraxial} approximation.  Let
\begin{equation}\label{new1}
E_x(x,y,z_0,t_0) \equiv E_0  e^{-\im  \omega t_0}, \quad
E_y(x,y,z_0,t_0) \equiv E_1  e^{-\im  \omega t_0},
\end{equation}
be the component of the complex paraxial electric field vector in
the $x$- and $y$-direction respectively, at the point $(x,y)$
located in the transverse plane $z = z_0$ at time $t_0$. If the
field is \emph{uniform} on the transverse plane, then $E_x$ and
$E_y$ will be, in fact, independent of $x$ and $y$ and  a complete
description of the field can be achieved in terms of a doublets
$\mathbf{E}$ of complex  variables (with possibly stochastic
fluctuations):
\begin{equation}\label{new2}
\mathbf{E} = \begin{pmatrix}
 E_{0} \\
 E_{1}
\end{pmatrix},
\end{equation}
where $E_0$ and $E_1$ are now complex-valued functions of $z_0$
and $t_0$ only. A complete study of the propagation of
$\mathbf{E}$ along $z$ can be found, e.g., in \cite{MandelBook},
however,  for our purposes the main result we need  is that
propagation through non-depolarizing media can be described by a
\emph{deterministic} Mueller (or Mueller-Jones) matrix  $M^J$,
while to describe the propagation of a light beam through a
depolarizing medium it is necessary to use a
\emph{non-deterministic} Mueller matrix $M$.
\subsubsection{Deterministic Mueller matrix $M^J$}
In a  wide-sense, a \emph{deterministic} linear scatterer as,
e.g., a quarter-wave plate, a rotator or a polarizer, is an
optical system which can be described by a $2 \times 2$ complex
Jones \cite{KligerBook} matrix
\begin{equation}\label{new3}
J = \begin{pmatrix}
 J_{00} &  J_{01}\\
 J_{10} &  J_{11}
\end{pmatrix}.
\end{equation}
With this we mean that if $\mathbf{E}$ and $\mathbf{E}'$ describe
the polarization state of the field immediately before and
immediately after the scatterer respectively, then they are
linearly related by the matrix $J$:
\begin{equation}\label{new4}
\mathbf{E}' = J \mathbf{E}.
\end{equation}
An alternative description  can be given in terms of the Stokes
parameters of the beam. To this end let $C$ be the covariance
matrix of the field defined as \cite{BornWolf}
\begin{equation}\label{new5}
C_{ij} = \la E_i E_j^* \ra, \qquad (i,j=0,1),
\end{equation}
where the brackets denote the statistical average over different
realizations of the random fluctuations of the field. Then the
four Stokes parameters $S_\mu$ $(\mu = 0, \ldots,3)$ of the beam
are defined as
\begin{equation}\label{new6}
S_\mu = \mathrm{Tr}\{ C \sigma_\mu\}, \qquad(\mu = 0, \ldots,3),
\end{equation}
were the symbol $\mathrm{Tr}\{ \cdot \}$ denote the trace
operation and the $\{\sigma_\mu \}$ are the normalized Pauli
matrices:
\begin{equation}\label{new7}
\begin{array}{lcl}
  \sigma_0 = \frac{1}{\sqrt{2}} \begin{pmatrix}
    1 & 0 \\
    0 & 1 \
  \end{pmatrix}, & \qquad &   \sigma_1 = \frac{1}{\sqrt{2}}\begin{pmatrix}
    0 & 1 \\
    1 & 0 \
  \end{pmatrix}, \\\\
    \sigma_2 = \frac{1}{\sqrt{2}}\begin{pmatrix}
    0 & -\im \\
    \im & 0 \
  \end{pmatrix}, & \qquad &  \sigma_3 = \frac{1}{\sqrt{2}} \begin{pmatrix}
    1 & 0 \\
    0 & -1 \
  \end{pmatrix}.
\end{array}
\end{equation}

Now, if with $S_\mu$ and $S_\mu'$ we denotes the Stokes parameters
of the beam before and  after the scatterer respectively, it is
easy to show that that they are linearly related by the
real-valued $4 \times 4$ Mueller-Jones matrix $M^J$ as
\begin{equation}\label{new8}
S_\mu' = M^J_{\mu \nu}S_\nu,
\end{equation}
where summation on repeated indices is understood and
\begin{equation}\label{new9}
M^J = \Lambda^\dagger (J \otimes J^*) \Lambda,
\end{equation}
where the symbol ``$\otimes$'' denotes the outer matrix product
and the \emph{unitary} matrix $\Lambda$ is defined as
\begin{equation}\label{new10}
\Lambda = \frac{1}{\sqrt{2}} \begin{pmatrix}
  1 & 0 & 0 & 1 \\
  0 & 1 & -\im & 0 \\
  0 & 1 & \im & 0 \\
  1 & 0 & 0 & -1
\end{pmatrix}.
\end{equation}
From the structure of $M^J$ follows that a deterministic medium
does not depolarize, that is $P_F(S) = P_F(S')$ where the degree
of polarization $P_F$ of the field is  defined as
\begin{equation}\label{new7b}
P_F(S) = \frac{\sqrt{S_1^2 + S_2^2 + S_3^2}}{S_0}.
\end{equation}
Let us conclude by noticing that  for deterministic media the two
descriptions in terms of $J$ or $M^J$ are completely equivalent in
the sense that the $16$ real elements of $M^J$ do not contain more
information than the $4$ complex elements of $J$.
\subsubsection{Non-deterministic Mueller matrix $M$}
A  \emph{non}-deterministic scatterer is, in a wide-sense, an
optical systems which \emph{cannot} be described by a
Mueller-Jones matrix. In this class fall all the depolarizing
optical system as, e.g., multi-mode optical fibers, particles
suspensions, etc..
It has been shown \cite{Kim87,Gil00} that  it is possible to
describe a non-deterministic optical system as an ensemble of
deterministic systems, in such a way that each realization
$\mathcal{E}$ in the  ensemble is characterized by a well-defined
Jones matrix $J(\mathcal{E})$ occurring with probability
$p_\mathcal{E} \geq 0$.  Then, the Mueller matrix $M$ of the
system can be written as
\begin{equation}\label{new11}
M = \Lambda^\dagger  (\overline{J \otimes J^*}) \Lambda
\end{equation}
where the $\overline{\mathrm{bar}}$ symbol  denotes the average
with respect to the ensemble representing the medium:
\begin{equation}\label{new12}
\overline{J \otimes J^*} = \sum_{\mathcal{E}} p_\mathcal{E}
J(\mathcal{E}) \otimes J^*(\mathcal{E}).
\end{equation}
At this point it is useful to introduce the auxiliary $4 \times 4
$ Hermitian matrix $H$ defined as
\begin{equation}\label{new12b}
H_{\mu \nu} = \overline{J_{ij}J_{kl}^*}, \qquad (\mu = 2i+j, \nu =
2k+l),
\end{equation}
which is, by definition, positive semidefinite, that is all its
eigenvalues $\{\lambda_0,\lambda_1,\lambda_2,\lambda_3 \}$ are
non-negative. Then, it is possible to show that
 the depolarization index $D_M$ and the polarization
entropy $E_M$ can be written as
\begin{equation}\label{eq:10}
D_M = \left[ \frac{1}{3}\left( 4 \sum_{\nu=0}^3 \lambda_\nu^2 - 1
\right) \right]^{1/2},
\end{equation}
\begin{equation}\label{eq:11}
E_M = -\sum_{\nu=0}^3 \lambda_\nu \log_4(\lambda_\nu ).
\end{equation}
From ref. \cite{aiello2} we know that $E_M$ is a multi-valued
function of $D_M$ and that this dependence determines some
physical bounds to \emph{any} polarization scattering process. The
function $E_M(D_M)$ shows thus character of universality. In the
next Section we shall confirm this theoretical prediction with
experimental results.
\subsection{Unpolarized light and depolarizing media}
In classical optics, a light beam appears to be depolarized when
its polarization direction varies rapidly with respect to other
degrees of freedom that are not measured during the experiment
(e.g. wavelength, time or position of the beam) \cite{Kliger}.
Moreover, depolarization occurs also when a single-mode
polarization input beam is coupled with a multi-mode (either
spectral or spatial) system as, e.g., an optical fiber.
%
%
%
%
In fact it is possible to identify two basic  depolarizing
processes, (a) one intrinsic to the medium,  and  (b) one due to
the measurement scheme. In the first case (a)  non-stationary
temporal fluctuations of the optical properties of the medium, for
instance due to the Brownian motion of suspended particles in a
liquid \cite{mckintosh}, cause depolarization even when a
single-mode detection scheme is employed (the time average
performed during the measurement is responsible for the
depolarization). On the other hand, type (b) stationary
depolarizers (i.e. glass fibers) do not fluctuate in time and
produce light depolarization only in the presence of a multi-mode
detection scheme. In this case it is simple to build explicitly
the ensemble of Mueller-Jones matrices representative of the
medium, that we
introduced in the previous subsection.\\
%
%
%
\noindent To this end, let us consider the case of a scattering
process in which a coupling between polarization and spatial modes
of the field occurs and a multi-mode detection scheme is employed.
This is, in fact, the case occurring for the optical fibers we
used.
The Mueller-Stokes formalism, is suitable for a single-mode
description of the field; however, it is possible to extend this
formalism to the case in which $N_\mathrm{in}$ spatial modes of
the field impinge on the scatterer, $N_\mathrm{out}$ leave from it
and $D$ modes are eventually detected. We make the assumption that
different spatial modes of the field are uncorrelated, that is we
do not consider interference phenomena which are ruled out by the
required linearity with respect to the intensities of the field
\cite{BornWolf}. Moreover, without loss of generality, we assume
$N_\mathrm{in}=N_\mathrm{out}=N$. Let $\mathbf{S} (j) \equiv \{
S_0(j), S_1(j),S_2(j), S_3(j)\}$ be a generic $4$-D Stokes vector
defined with respect to the mode $j$, where $j \in \{1, \ldots, N
\}$. For a $N$-mode field we have a collection of $N$ of these
$4$-D Stokes vectors that  we can arrange in a single $4N$-D
``super'' vector $ \mathbb{S} = \{\mathbf{S}(1), \ldots,
\mathbf{S}(N) \}$. When the $N$-mode light beam undergoes a
polarization-sensitive scattering, then, in general, the Stokes
vectors $\{ \mathbf{S}_\mathrm{in}(j) \}$ of the input beam are
related to the set of vectors $\{ \mathbf{S}_\mathrm{out}(j) \}$
of the output beam by:
\begin{equation}\label{eq:2}
\mathbf{S}_{out}(j) = \sum_{j_0 =1}^N M^J(j,j_0)
\mathbf{S}_{in}(j_0), \qquad (j = 1, \ldots, N ),
 \end{equation}
where $M^J(j,j_0)$ is the $4 \times 4$ Mueller-Jones matrix that
describes the scattering from the input mode ${j_0}$ to the output
mode ${j}$. If we introduce a ``super'' $4N \times 4N$ Mueller
matrix defined as
\begin{equation}\label{eq:3}
\mathbb{M} \equiv \begin{pmatrix}
  M^J(1,1) & \dots &  M^J(1,N) \\
  \vdots & \ddots & \vdots \\
   M^J( N,1) & \dots &  M^J(N,N)
\end{pmatrix},
 \end{equation}
where each block $M^J(j_1,j_2)$ is a $4 \times 4$ Mueller-Jones
matrix, then we can rewrite Eq. (\ref{eq:2}) in a compact form as
\begin{equation}\label{eq:4}
\mathbb{S}_{out} = \mathbb{M}\cdot \mathbb{S}_{in}.
 \end{equation}
After the scattering process  took place, we have to detect its
products. We recently showed \cite{Aiello04_1} that when $D < N$
modes of the field are detected, a mode-insensitive polarization
analyzer, put in front of a bucket-detector, can be described by a
$4N \times 4N$ block-diagonal matrix $\mathbb{A}$:
\begin{equation}\label{eq:5}
\mathbb{A} \equiv \begin{pmatrix}
  A(1) &  &  &  &\\
   &  \ddots & & &\\
    &  &  A(D) & & \\
       &  &  & \mathbf{0} &
\end{pmatrix},
 \end{equation}
where $A(j), \; (j = 1,\ldots,D)$ are  $4 \times 4$ real-valued
positive semi-definite matrices (in fact, projectors), and
$\mathbf{0}$ is a null $(N-D) \times (N-D)$ matrix. In the
paraxial limit ($D<<N$) each $A(j)$ reduces to the $4\times4$
identity.
 Then, the
polarization state of the scattered beam after the analyzer, is
described by the super Stokes vector $\mathbb{S}_D$ given by
\begin{equation}\label{eq:6}
\mathbb{S}_D = \mathbb{A} \cdot \mathbb{M} \cdot \mathbb{S}_{in}.
 \end{equation}
Finally, because of the mode-insensitive detection,  the sum over
all the detected modes reduces  the number of degrees of freedom
of the field from $4N$ to $4$, producing the \emph{detected} 4-D
Stokes vector $\mathbf{S}_{D}$
\begin{equation}\label{eq:7}
\mathbf{S}_D = \displaystyle{\sum_{j=1}^{N} \mathbf{S}_{D}(j)}
%
%
= M_{\mathrm{\emph{eff}}}\mathbf{S}_{in}(1),
 \end{equation}
where  we have assumed that the input light beam is prepared in
the single mode $j_0=1$, so that $\mathbf{S}_{in}(j_0) =
\delta_{j_0 1} \mathbf{S}_{in}(1) $ and with $M_\mathrm{eff}$ we
have denoted an effective $4 \times 4$ Mueller matrix defined as
\begin{equation}\label{eq:8}
M_{\mathrm{\emph{eff}}} = \sum_{j=1}^{D} A(j) M^J(j,1),
 \end{equation}
which is written as a sum of $D$ Mueller-Jones matrices. It is
important to notice that while the product of Mueller-Jones
matrices is still a Mueller-Jones matrix (in physical terms: a
cascade of non-depolarizing optical elements is still a
non-depolarizing optical system), a sum, in general, is not. This
causes depolarization.
Moreover, since the ``matrix coefficients'' $A(j)$ are
non-negative, the matrix $M_{\rm{eff}}$ in Eq. (\ref{eq:8}) is an
explicit version of the Mueller matrix written in Eq.
(\ref{new11}). Then we have shown, by an explicit derivation, how
to build the statistical ensemble representing the depolarizing
medium, for this particular case.
\section{DEPOLARIZATION EXPERIMENTS}
\bigskip
\subsection{Experimental scheme for polarization tomography}

\begin{figure}
\includegraphics[angle=0,width=15 truecm]{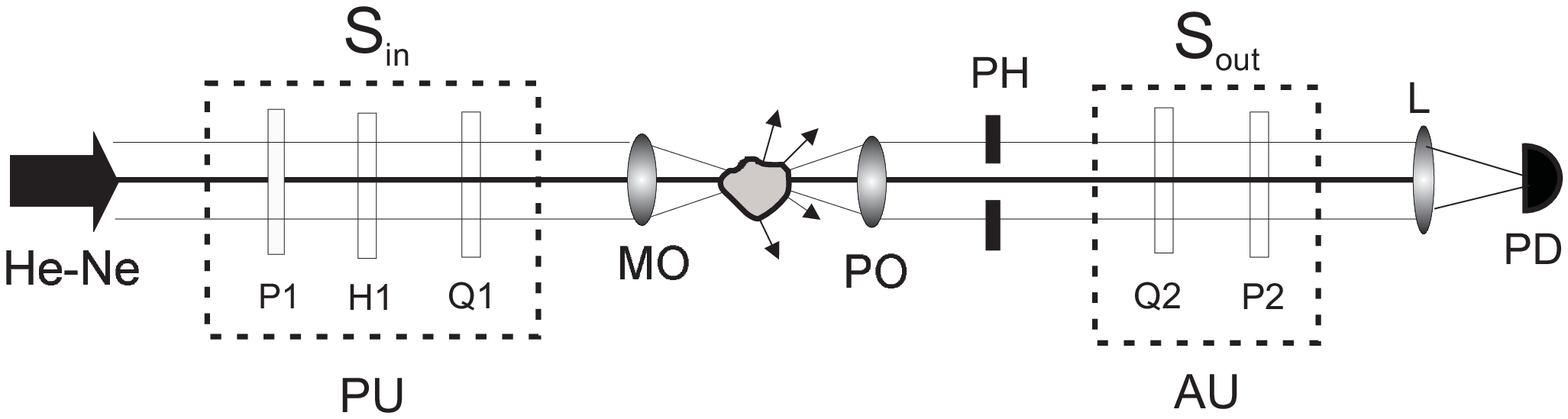}
\caption{\label{fig:1} Schematic of the polarization tomography
set-up. For details see text.}
\end{figure}
In order to measure the effective Mueller matrix $M_{\rm{eff}}$
and thus the index of depolarization $D_M$ and the entropy $E_M$
of a scattering medium, it is straightforward to follow a
tomography procedure: The light to be scattered by the sample is
successively prepared in the four polarization basis states of
linear ($V, H, +45^\circ$) and circular ($RHC$) polarization,
which are represented by four independent input Stokes vectors
$\mathbf{S}_{\rm in}$. For each of these input fields the
corresponding Stokes vector $\mathbf{S}_{\rm out}$, that
represents the output field, is obtained by measuring the
intensities of the scattered light in the same four polarization
basis states. This procedure provides the $4\times4$ independent
parameters required to determine the 16 elements $M_{\mu\nu}$ of
the Mueller matrix from Eq.(\ref{new8}). Note that we actually
employ two additional polarization basis states ($-45^\circ$,
$LHC$) in our experiments and perform $6\times6$ measurement,
which allows us to reduce experimental errors by averaging within
the over-complete data set.

The experimental scheme is illustrated in Fig.~\ref{fig:1}. The
light source is a power-stabilized He-Ne laser at 633~nm
wavelength. The input field is prepared by the polarizer unit
(PU), consisting of a fixed polarizer (P1), a half-wave plate
(H1), and a quarter-wave plate (Q1). A microscope objective (MO,
$\times50/0.55$) couples the light into the sample. The scattered
light is collimated by a standard photographic objective (PO,
$50$~mm$/1.9$), followed by an adjustable pinhole (PH) that
defines the amount of transverse spatial average to be performed
in the light detection. The analyzer unit (AU) consists of a
quarter-wave plate (Q2) and a polarizer (P2). Together with a
focusing lens (L) and a photodiode (PD), it probes the
polarization state of the scattered output field.
As an estimation of the systematic error of the set-up, mainly due
to imperfections of the used retarders, we measured the Mueller
matrix of air ({i.e.}~the identity matrix) and of well-known
deterministic optical elements such as wave-plates. In all these
cases, we found the deviations from the theoretically predicted
matrix elements limited to $|\Delta M_{\rm \mu\nu}|\leq\pm 0.04$.

\subsection{Collection of scattering media}

The various scattering media we investigated can be divided into
(a) \emph{non-stationary} samples where, {e.g.}, Brownian motion
induces temporal fluctuations within the detection integration
time, and (b) \emph{stationary} samples without such fluctuations,
most notably multi-mode polymer  and glass  optical fibers. More
specifically, we chose our scatterers from:

\begin{itemize}
  \item[(a)] Non-stationary media:
    \begin{itemize}
       \item polystyrene microspheres ($2~\mu$m dia., suspended in
      water,
            \emph{Duke Scientific Co.}, USA).
       \item diluted milk;
       \end{itemize}
  \item[(b)] Stationary media:
    \begin{itemize}
      \item Zenith$^{\rm TM}$ polymer sheet diffusers ($100~\mu$m
      thick,
            \emph{SphereOptics Hoffman {GmbH}}, Germany);
      \item holographic light shaping diffusers ($0.5^\circ$, $1^\circ$,
            $5^\circ$, and $10^\circ$ scattering angle,
            \emph{Physical Optics Co.}, USA);
      \item quartz/silica wedge depolarizers and quartz Lyot
            depolarizers, \emph{Halbo Optics}, UK);
      \item step-index polymer optical fiber (NA=0.55, core dias.~$250~\mu$m, $500~\mu$m,
            $750~\mu$m \emph{ESKA CK} type, \emph{Mitsubishi Rayon}, Japan);
      \item step-index glass optical  fiber (NA=0.48, core dias.~$200~\mu$m, $400~\mu$m,
            $600~\mu$m \emph{FT-x-URT} type, distributed by \emph{Thorlabs, Inc.}, USA);
      \item step-index  glass optical fiber (NA=0.22, core dia.~$50~\mu$m,
            \emph{ASF50} type, distributed by \emph{Thorlabs, Inc.}, USA);
      \item graded-index glass optical fiber (NA=0.27, core dia.~$62,5~\mu$m,
            \emph{GIF625} type, distributed by \emph{Thorlabs, Inc.}, USA).
    \end{itemize}
\end{itemize}

\subsection{Experimental results}
\bigskip

\begin{figure}[h!]
\includegraphics[angle=0,width=15 truecm]{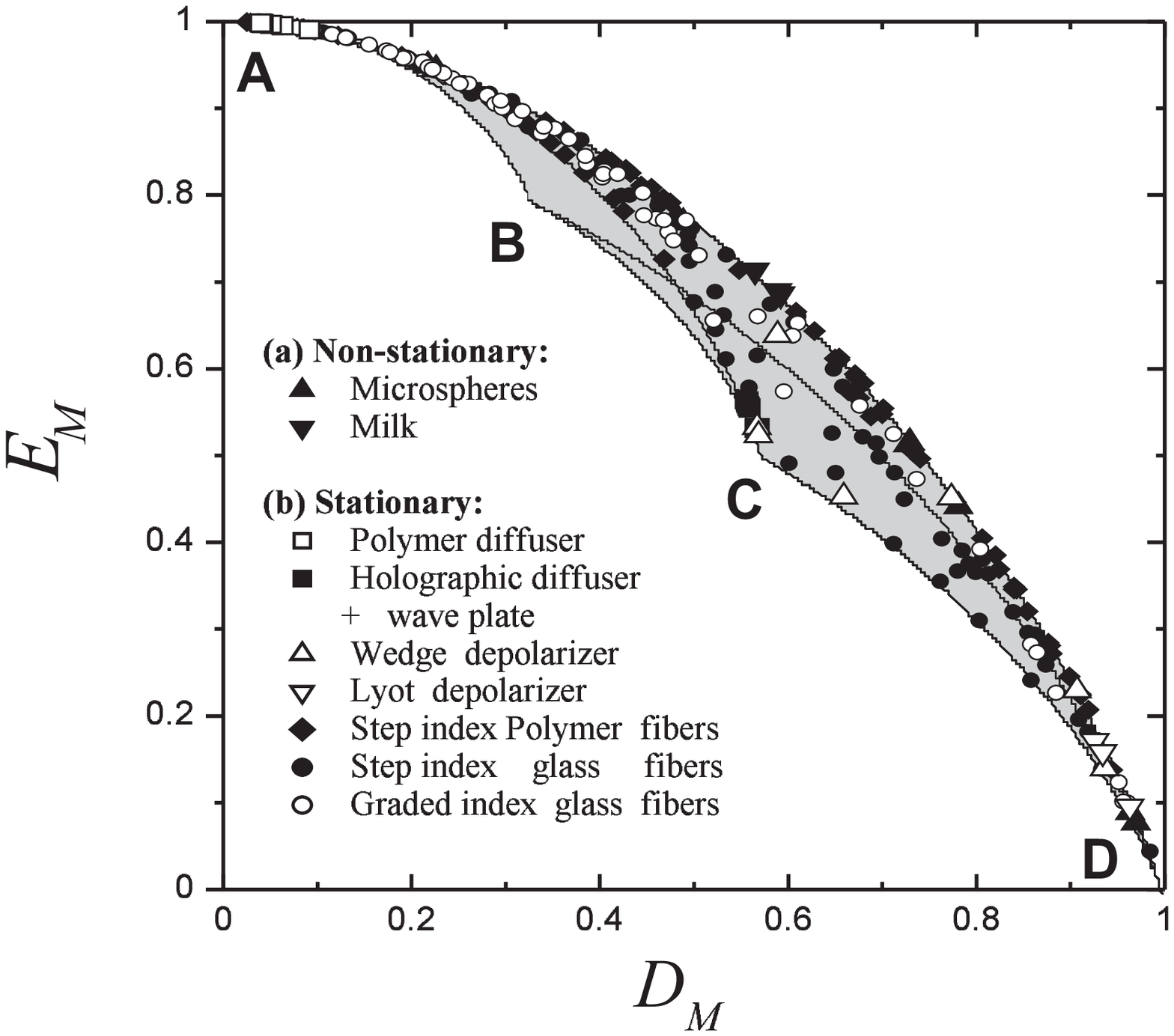}
\caption{\label{fig:2} Measured entropy $(E_{M})$ vs. index of
depolarization $(D_{M})$ for (a) non-stationary and (b) stationary
scattering media. The maximal possible parameter range in the
$(D_{M}, E_{M})$ plane is indicated by the grey-shaded area. Lines
correspond to analytical boundaries predicted by theory. Cuspidal
points are given by $\mathbf{A}=(0,1)$,
$\mathbf{B}=(1/3,\log_{4}3)$, $\mathbf{C}=(1/\sqrt3,1/2)$, and
$\mathbf{D}=(1,0)$ \cite{aiello2}.}
\end{figure}

\begin{figure}[h]
\includegraphics[angle=0,width=15truecm]{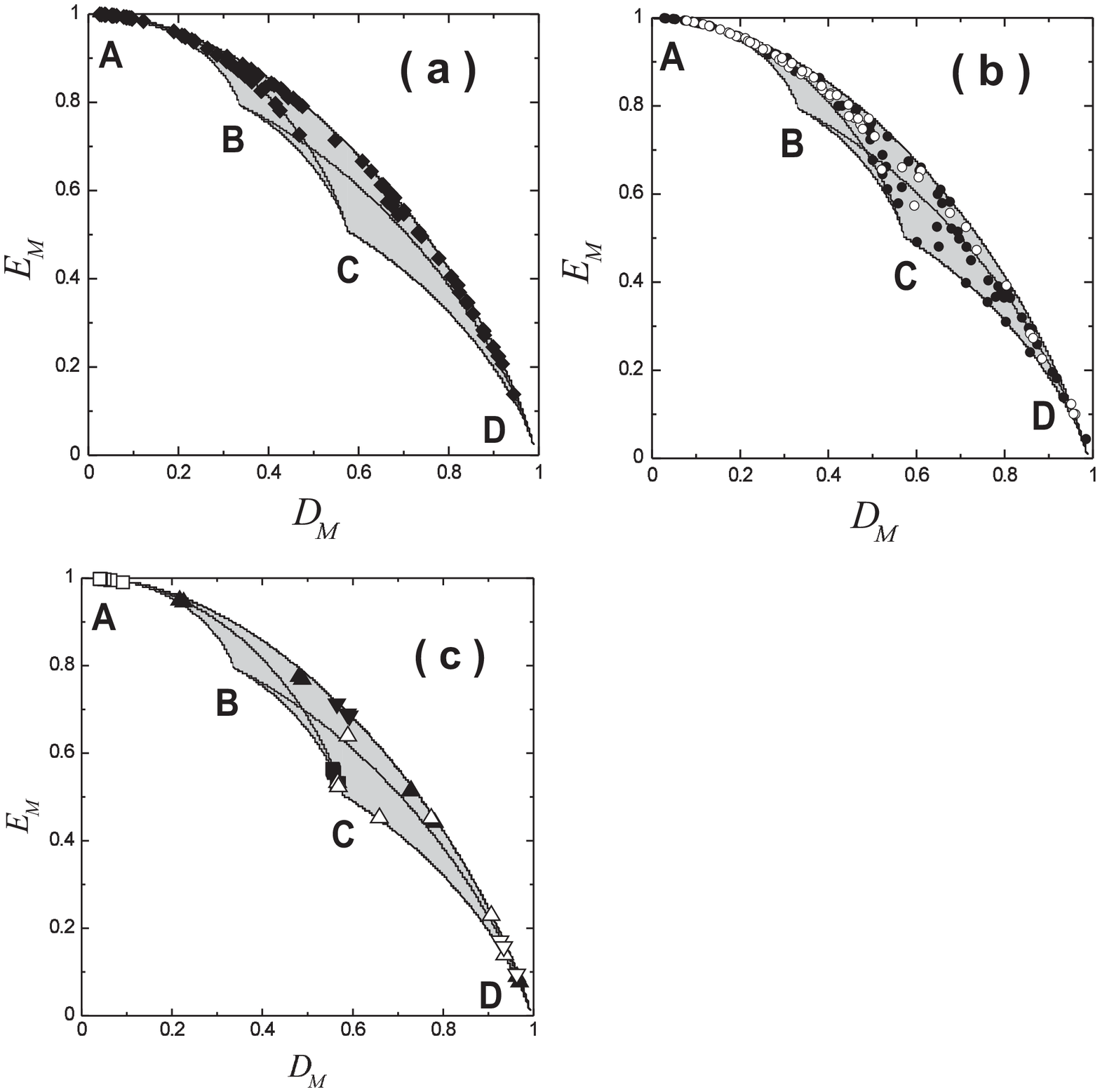}
\caption{\label{fig:3} Measured entropy $(E_{M})$ vs. index of
depolarization $(D_{M})$ for (a) step-index polymer optical fibers
($\blacklozenge$); (b) step-index glass optical fibers
($\bullet$), graded-index glass optical fibers ($\circ$); (c)
microspheres ($\blacktriangle$), milk ($\blacktriangledown$),
polymer diffuser ($\square$), holographic diffusers with wave
plate ($\blacksquare$), wedge depolarizers ($\vartriangle$), and
Lyot depolarizers ($\triangledown$). The analytical boundaries are
indicated by the continuous lines.}
\end{figure}

For a large collection of different samples, Fig.~\ref{fig:2}
shows the measured polarization entropy $E_{M}$ vs. the
corresponding index of depolarization $D_{M}$. The black lines
represent the calculated analytical boundaries in the
$(E_{M},D_{M})$ plane, whose functional dependence $E_{M}(D_{M})$
was derived in Ref.~\cite{aiello2}. These boundaries provide
universal constraints to the possible values $(E_{M}, D_{M})$ for
\emph{any} physical scattering system, that is, the range of
admissible values is restricted to the rather limited grey-shaded
area within the boundaries. As it is apparent from the
experimental data, our choice of samples allowed us to widely fill
in the range of values $(E_{M},D_{M})$, in good agreement with the
prediction from Ref.~\cite{aiello2}. For rather different
scattering media, we observed similar values of the pairs
$(E_{M},D_{M})$, which display  the universality in this
quantitative description of the depolarizing properties. We found
samples throughout the full range of values in entropy and
depolarizing power, $0\leq E_{M}, D_{M}\leq 1$. However, note that
the region below the curve connecting the points {\bf A} and {\bf
C} in the $(E_{M},D_{M})$ plane is not covered by any data so far.
Work is in progress to investigate this peculiarity.\\
Several scatterer-specific tuning parameters allowed us to realize
this wide range of depolarizing systems and to reveal details of
the depolarizing properties for the various media, as will be
discussed in the following subsection.\\
The most versatile scatterers used to acquire data in the
$(E_{M},D_{M})$ plane, were the multi-mode optical fibers. For
them, the depolarization is caused by multiple reflections within
the cylindrical light-guide together with mode mixing. We selected
fibers of various lengths that displayed the full range between
non-depolarizing and completely depolarizing properties. Fibers
shorter than about 2~cm showed negligible depolarization
($D_{M}\approx 1$). In the case of the glass fibers complete
depolarization ($D_{M}\approx 0$) was observed for lengths of
$\approx 5~$m, whereas in the case of the polymer fibers  this was
achieved already for lengths of only $\approx 50~$cm. The reason
is, presumably, the significant Rayleigh scattering at density
fluctuations in the
polymer material \cite{POF}.\\
In our experimental scheme, the aperture of the pinhole (PH)
defines the region of spatial of average in the scattered light
detection (see Fig.~\ref{fig:1}). By choosing the pinhole diameter
between 2~mm and 13~mm, we realized scattering systems which are
described by different effective Mueller matrices $M_{\rm{eff}}$.
In fact, a small pinhole, corresponding to an average over a small
set of  modes $j$ in Eq.~(\ref{eq:8}), leads to a large index of
depolarization. However, due to the huge optical mode volume in
our fibers, the pinhole adjustment allowed only for small changes,
within $|\Delta D_{M}|\leq\pm 0.05$. With a lower limit of 2~mm in
the pinhole diameter, special care was taken to select a
sufficiently large number of speckles in the scattered output
field. (The step-index glass fiber with $50~\mu$m core showed the
largest speckles of about $500~\mu$m FWHM.) This is necessary in
order to average out interference effects, generated by the
coherent source (He-Ne laser),  so that the assumption of
uncorrelated modes in the derivation of $M_{\rm{eff}}$ holds
\cite{unphysical}.

\subsection{Discussion}

In Fig.~\ref{fig:3}, we separately show the results for (a)
step-index polymer fibers, (b) step-index and graded-index glass
fibers, and (c) other scattering media. For the polymer fibers it
is apparent that most of the data fall on the upper curve
connecting the points ${\bf A}$ and ${\bf D}$. This curve
corresponds to Mueller matrices that have an associated operator
$H$ with its four eigenvalues of the degenerate form \{$\lambda,
\mu, \mu, \mu$\}. This degeneracy can be associated with
\emph{isotropic} depolarizers, which is obviously a good
description for the polymer fibers. Contrarily, in
Fig.~\ref{fig:3}(b) (glass fibers), we fill in also the allowed
$(E_M, D_M)$ domain below the isotropy curve. These domains
correspond to \emph{anisotropic} media, the anisotropy being
supposedly due to stress-induced birefringence in the glass
fibers. It was actually this birefringence which we used as an
additional tuning parameter accounting for changes of a few
percent in the index of depolarization. The data obtained for long
fiber samples, both in polymer and glass,  are close to the
cuspidal point {\bf A}~($\lambda=\mu=1/4$), which corresponds to
total depolarizers. Contrarily, the data for very short fiber
samples are close to the cuspidal point {\bf D}~($\lambda=1,
\mu=0$), which corresponds to a deterministic (i.e.
non-depolarizing) system. \\
In case of watery suspensions such as
milk and microspheres, we observed purely isotropic
depolarization, {i.e.}~all data are found on the isotropic curve,
similar to the polymer optical fibers (see Fig.~\ref{fig:3}(c)).
We adjusted the depolarizing power by varying the concentration of
the scatterers. The $100~\mu$m thick Zenith$^{\rm TM}$ polymer
diffuser sheet was found to be almost completely depolarizing,
whereas the holographic diffusers, when used alone, did not cause
any significant depolarization. The latter effect is due to the
absence of {\it multiple} scattering in the transmission of light
through these surface-optical elements. In combination with a
subsequent wave-plate we could, however, couple the scattered
light angular spectrum to the polarization degrees of freedom, and
thus achieve
depolarization.\\
Finally, the data for standard wedge and Lyot depolarizers are
also shown in Fig.~\ref{fig:3}(c). Wedge depolarizers are designed
to completely depolarize a well defined linear input polarization,
whereas our tomographic measurement procedure represents an
average over all independent input polarizations. This results in
a non-zero index of depolarization. The Lyot depolarizer is,
within the experimental error, non-depolarizing since it is
designed to depolarize a broadband light source while we operated
with a monochromatic laser source.

\section{CONCLUSIONS}

By means of polarization tomography we have experimentally
characterized the depolarizing properties for a large set of
scattering optical media. We describe these media with both the
index of depolarization $D_{M}$ and  the polarization entropy
$E_{M}$ that is added by the medium to the scattered field. These
quantities are derived from a measured \emph{effective} Mueller
matrix, which we formally introduce in the description of
scattering systems subject to a multi-mode detection, as is the
case of our experimental configuration. The set of studied media
ranges from non-stationary scatterers such as milk and polystyrene
microspheres to stationary scatterers such as multi-mode optical
fibers, diffusers, and standard wedge depolarizers.\\
In Ref.~\cite{aiello2} a universality was predicted for the
possible values of $E_{M}$ and $D_{M}$, these values being
restricted to a limited domain  described by a set of analytical
boundaries. The collected experimental data for our scatterers
fill in this domain almost completely and give evidence of the
predicted depolarization universality in light scattering. A
certain range of the predicted $(E_{M},D_{M})$ values is, however,
not covered by the scatterers we investigated so far. Work on this
issue is currently under progress.
\\Furthermore, the quantities $D_{M}$ and $E_{M}$ provide insights
into the particular depolarization mechanisms of the various
media, as well as a consistency check for the measured data (see
\cite{unphysical}), and may provide a useful classification of
optical scatterers for quantum applications, where depolarization
stands for decoherence \cite{gisin}. In this spirit, an extension
to twin photon quantum scattering experiments is relatively
straight-forward. Work along this line is also under progress
in our group.\\

We have greatly benefited from many discussions with  Martin van
Exter and with  Eric Eliel, who are acknowledged. This project is
part of the program of FOM and is also supported by the EU under
the IST-ATESIT contract.


\begin{thebibliography}{ab22}
\expandafter\ifx\csname
natexlab\endcsname\relax\def\natexlab#1{#1}\fi
\expandafter\ifx\csname bibnamefont\endcsname\relax
  \def\bibnamefont#1{#1}\fi
\expandafter\ifx\csname bibfnamefont\endcsname\relax
  \def\bibfnamefont#1{#1}\fi
\expandafter\ifx\csname citenamefont\endcsname\relax
  \def\citenamefont#1{#1}\fi
\expandafter\ifx\csname url\endcsname\relax
  \def\url#1{\texttt{#1}}\fi
\expandafter\ifx\csname
urlprefix\endcsname\relax\def\urlprefix{URL }\fi
\providecommand{\bibinfo}[2]{#2}
\providecommand{\eprint}[2][]{\url{#2}}

\bibitem[{DOP()}]{DOP}
\bibinfo{note}{
J. F. de Boer, T. E. Milner, M. J. C. van Gemert, and J. S.
Nelson, Opt. Lett. {\bf 22}, 934 (1997);
%
J. M. Bueno and P. Artal, Opt. Lett. {\bf 24}, 64 (1999);
%
A.H. Hielscher {\it et al.}, Opt. Expr. {\bf 1}, 441 (1997); B.
%
Laude-Boulesteix, A. De Martino, B. Dr\'{e}villon, and L.
Schwartz, Appl. Opt {\bf 43}, 2824 (2004);}

\bibitem[{\citenamefont{Kliger et~al.}(1990)\citenamefont{Kliger, Lewis, and
  Randall}}]{KligerBook}
\bibinfo{author}{\bibfnamefont{D.~S.} \bibnamefont{Kliger}},
  \bibinfo{author}{\bibfnamefont{J.~W.} \bibnamefont{Lewis}}, \bibnamefont{and}
  \bibinfo{author}{\bibfnamefont{C.~E.} \bibnamefont{Randall}},
  \emph{\bibinfo{title}{Polarized Light in Optics and Spectroscopy}}
  (\bibinfo{publisher}{Academic Press, Inc.}, \bibinfo{year}{1990}).

\bibitem[{\citenamefont{Gil and Bernabeu}(1986)}]{Gil86}
\bibinfo{author}{\bibfnamefont{J.~J.} \bibnamefont{Gil}} \bibnamefont{and}
  \bibinfo{author}{\bibfnamefont{E.}~\bibnamefont{Bernabeu}},
  \bibinfo{journal}{Optica Acta} \textbf{\bibinfo{volume}{33}},
  \bibinfo{pages}{185} (\bibinfo{year}{1986}).

\bibitem[{\citenamefont{Roy-Brehonnet and Jeune}(1997)}]{LeRoy}
\bibinfo{author}{\bibfnamefont{F.} \bibnamefont{Le Roy-Brehonnet}}
  \bibnamefont{and} \bibinfo{author}{\bibfnamefont{B.} \bibnamefont{Le Jeune}},
  \bibinfo{journal}{Prog. Quant. Electr.} \textbf{\bibinfo{volume}{21}},
  \bibinfo{pages}{109} (\bibinfo{year}{1997}).

\bibitem{aiello2}A. Aiello, J. P. Woerdman, submitted Phys. Rev. Lett.,
arXiv:quant-ph/0407234.

\bibitem[{\citenamefont{{E. Wolf}}(2003)}]{Wolf03}
\bibinfo{author}{\bibnamefont{{E. Wolf}}}, \bibinfo{journal}{Phys. Lett. A} \textbf{\bibinfo{volume}{312}},
  \bibinfo{pages}{263} (\bibinfo{year}{2003}).

\bibitem{gisin}M Legr\'{e}, M. Wegm\"{u}ller and N. Gisin, Phys. Rev. Lett. \textbf{91},
167902 (2003).

\bibitem[{\citenamefont{Mandel and Wolf}(1995)}]{MandelBook}
\bibinfo{author}{\bibfnamefont{L.}~\bibnamefont{Mandel}} \bibnamefont{and}
  \bibinfo{author}{\bibfnamefont{E.}~\bibnamefont{Wolf}},
  \emph{\bibinfo{title}{Optical Coherence and Quantum Optics}}
  (\bibinfo{publisher}{Cambridge University Press}, \bibinfo{year}{1995}),
  \bibinfo{edition}{1st} ed.

\bibitem[{\citenamefont{Born and Wolf}(1984)}]{BornWolf}
\bibinfo{author}{\bibfnamefont{M.}~\bibnamefont{Born}} \bibnamefont{and}
  \bibinfo{author}{\bibfnamefont{E.}~\bibnamefont{Wolf}},
  \emph{\bibinfo{title}{Principles of Optics}} (\bibinfo{publisher}{Pergamon
  Press}, \bibinfo{year}{1984}), \bibinfo{edition}{sixth} ed.

\bibitem{Gil00}
\bibinfo{author}{J. J. Gil}, \bibinfo{journal}{J. Opt. Soc. Am. A}
 \textbf{\bibinfo{volume}{17}}, \bibinfo{pages}{328}
  (\bibinfo{year}{2000}).

\bibitem[{\citenamefont{Kim et~al.}(1987)\citenamefont{Kim, Mandel, and
  Wolf}}]{Kim87}
\bibinfo{author}{\bibfnamefont{K.}~\bibnamefont{Kim}},
  \bibinfo{author}{\bibfnamefont{L.}~\bibnamefont{Mandel}}, \bibnamefont{and}
  \bibinfo{author}{\bibfnamefont{E.}~\bibnamefont{Wolf}}, \bibinfo{journal}{J.
  Opt. Soc. Am. A} \textbf{\bibinfo{volume}{4}}, \bibinfo{pages}{433}
  (\bibinfo{year}{1987}).

\bibitem{Kliger}D. S. Kliger, J. W. Lewis, and C. E. Randall,
\textit{Polarized light in optics and spectroscopy}, Academic
Press Inc. (1990).

\bibitem{mckintosh} F.C. MacKintosh, J.X. Zhu, D.J. Pine and D.A.
Weitz, Phys. Rev. B (RC), \textbf{40}, 9342 (1989).

\bibitem[{\citenamefont{Aiello and Woerdman}(2004)}]{Aiello04_1}
\bibinfo{author}{\bibfnamefont{A.}~\bibnamefont{Aiello}} \bibnamefont{and}
  \bibinfo{author}{\bibfnamefont{J.~P.} \bibnamefont{Woerdman}},
  \bibinfo{journal}{Phys. Rev. A} \textbf{\bibinfo{volume}{70}},
  \bibinfo{pages}{023808} (\bibinfo{year}{2004}), quant-ph/0404029.

\bibitem{POF}
For a review on POF see, {e.g.}, {J.} Zubia and {J.} Arrue, Opt.
Fib. Technol. {\bf 7}, 101 (2001).

\bibitem{unphysical}
When measuring with a pinhole in size similar with the speckles,
we derived many negative eigenvalues $\lambda_{i}$. The resulting
{\it complex-valued} entropies $E_{M}$ are not even mathematically
well defined, see Eq.~(\ref{eq:11}).

\end{thebibliography}
\end{document}